\documentclass[twoside,a4paper,11pt]{sea9}
\usepackage{graphicx}
\usepackage{hyperref}
\usepackage{movie15}
\newcommand{\hh}{H\,{\footnotesize II}~}

\newcommand{\nii}{[N\,{\footnotesize II}]}
\newcommand{\sii}{[S\,{\footnotesize II}]}

\newcommand{\oi}{[O\,{\footnotesize I}]}
\newcommand{\oii}{[O\,{\footnotesize II}]}
\newcommand{\oiii}{[O\,{\footnotesize III}]}

\newcommand{\ha}{H$\alpha$}

\begin{document}
\pagenumbering{arabic}
\pagestyle{myheadings}
\thispagestyle{empty}
{\flushleft\includegraphics[width=\textwidth,bb=58 650 590 680]{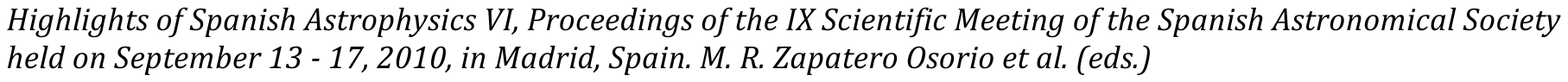}}
\vspace*{0.2cm}
\begin{flushleft}
{\bf {\LARGE
%
Integral Field Spectroscopy surveys of nearby spiral and U-LIRG galaxies
%
}\\
\vspace*{1cm}
%
F.~F. Rosales-Ortega$^{1,2}$,
S. Arribas$^{1}$
and
R.~C. Kennicutt$^{3}$ 
%
}\\
\vspace*{0.5cm}
%
{\small 
$^{1}$
Consejo Superior de Investigaciones Cient{\'i}ficas, INTA-CAB, Ctra de Ajalvir
km 4, 28850, Torrej{\'o}n de Ardoz, Madrid, Spain.\\
$^{2}$
Departamento de F{\'i}sica Te{\'o}rica, Universidad Aut\'onoma de Madrid,
28049 Madrid, Spain.\\
$^{3}$
Institute of Astronomy, University of Cambridge, Madingley Road, Cambridge CB3
0HA, UK.
}
%
\end{flushleft}
%
\markboth{
IFS surveys of nearby spiral and (U)LIRG galaxies
}{ 
%
F.~F. Rosales-Ortega et al.
%
}
\thispagestyle{empty}
\vspace*{0.4cm}
\begin{minipage}[l]{0.09\textwidth}
\ 
\end{minipage}
\begin{minipage}[r]{0.9\textwidth}
\vspace{1cm}
\section*{Abstract}{\small

Here we describe the observations and preliminary results of the gas-phase
analysis based on two ongoing, wide-field Integral Field Spectroscopy (IFS)
surveys: the PPAK IFS Nearby Galaxies Survey (PINGS), targeting disc galaxies; and the
VIMOS-IFU observations of low-z (Ultra)Luminous Infrared Galaxies (U-LIRGs), the
local counterpart of massive, dusty high-z star-forming galaxies. 
We describe how these observations are allowing to discover and characterise
abundance differentials between galactic substructures and differences depending
on the morphologically/dynamically distinct types of objects, which in turn will
allow us to interpret the gas-phase abundances of analogue high-z systems.

\normalsize}
\end{minipage}
%
%

\section{Introduction \label{intro}}

The nebular emission arising from bright \hh regions within actively
star-forming galaxies has played an important role in our understanding of the
chemical evolution in the universe.
Metals are a fundamental parameter for cooling mechanisms in the intergalactic
and interstellar medium (ISM), star formation, stellar physics, and planet
formation. Measuring the chemical abundances in individual galaxies
and galactic substructures, over a wide range of redshifts, is a crucial step to
understanding the chemical evolution and nucleosynthesis at different epochs, since
the heavy atomic nuclei trace the evolution of past and current stellar
generations.

This evolution is dictated by a complex array of parameters, including
the local initial gas composition, star formation history (SFH), gas infall and
outflows, radial transport and mixing of gas within discs, stellar yields, and
the initial mass function. Although it is difficult to disentangle the effects of
the various contributors, determinations of current elemental abundances constrain
the possible evolutionary histories of the existing stars and galaxies, and the
interaction of galaxies with the intergalactic medium. The details of such a
complex mechanism are still observationally not well established and
theoretically not well developed, and threaten our understanding of galaxy
evolution from the early universe to present day.


Rest-frame optical nebular emission lines have been --historically-- the
main tool at our disposal to the direct measurement of the gas-phase abundance at
discrete spatial positions in low-z galaxies (e.g. Pagel et al. \cite{pagel}, Garnett
\cite{gar}). The relative strengths of these lines provide a diagnostic tool to
identify both the abundance of heavy elements and the spectral shape of the
ionizing flux distribution.
The relevance of the study of the ISM in the local Universe cannot be
underestimated, since the diagnostic tools developed to determine the abundance
of heavy elements in the nearby Universe constitute the basis of the methods
employed to derive abundances (and their relations with global galaxy
parameters) in high-z galaxies (e.g. Nagao et al. \cite{nagao}, Maiolino et
al. \cite{maio}), objects that are typically characterised spectroscopically by
their emission lines.

\begin{figure}
\center
\includegraphics[height=6cm]{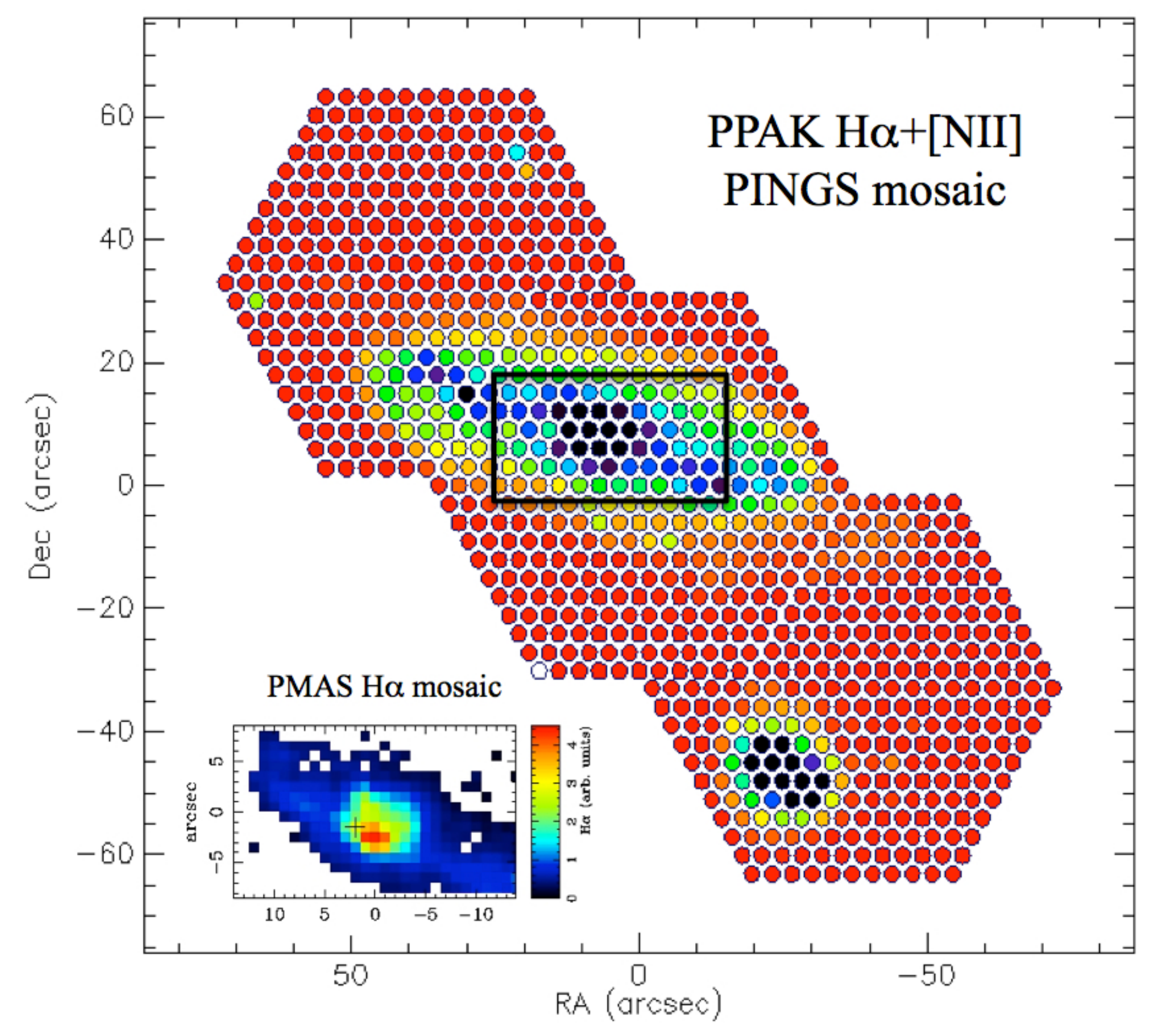} ~
\includegraphics[height=6cm]{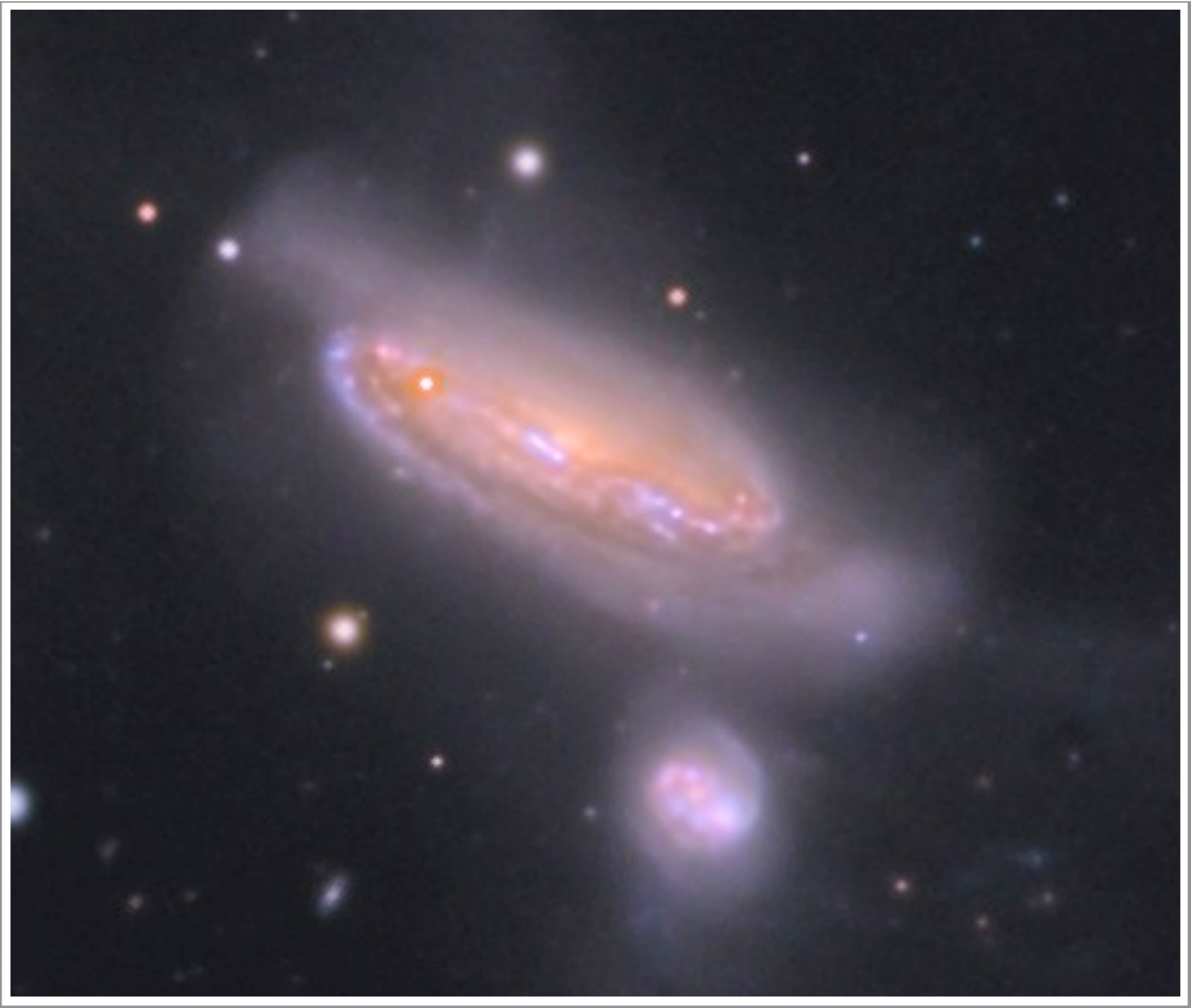} 
\caption{\footnotesize
  Example of an IFS mosaic observed by PINGS.
  {\em Left}: H$\alpha$+\nii\ ``narrow band image'' of the PINGS mosaic
  containing the interactive galaxies NGC\,7770 and NGC\,7771. The
  inner-bottom panel shows a PMAS H$\alpha$ emission line map of the centre of
  NGC\,7771 (black rectangle in the large mosaic) after Alonso-Herrero et
  al. \cite{alonso}.
   {\em Right}: Colour composite image of the NGC\,7770, NGC\,7771 interactive
  system (Image by Kent Biggs).
  \label{fig1} 
}
\end{figure}

There have been several investigations of possible relationships between the
global abundance properties of galaxies and their structural characteristics.
However, the reduced number of galaxies studied in detail and the small number of
\hh regions studied per galaxy preclude definitive conclusions. Until recently,
most of these measurements were made with single-aperture or long-slit
spectrographs, resulting in samples of typically a dozen or fewer \hh regions
per galaxy or single spectra of large surveys samples like the Sloan Digital Sky
Survey (SDSS). However, galaxies are complex systems not fully represented by a
single spectrum or just broad band colours. Disk and spheroidal components are
structurally and dynamically different entities with different SFH and chemical
evolution, fact that has not been properly addressed until now.

The advent of Multi-Object Spectrometers (MOS) and Integral Field Spectroscopy
(IFS) instruments with large fields of view (FOV) now offers us the opportunity
to undertake a new generation of emission-line surveys, based on samples of
scores to hundreds of \hh regions and full two-dimensional (2D) coverage of the
disks of nearby galaxies. 
This novel approach is being implemented in a new series of ambitious IFS
surveys around the world, most notably the SAURON instrument (Bacon et
al. \cite{bacon}), which has been extensively used to survey early type
galaxies; 
the VENGA project (Blanc et al. \cite{blanc}) which plans to use the VIRUS-P
instrument to carry out a spectroscopic mapping in a sample of nearby galaxies;
the CALIFA survey (see contribution by S. S\'anchez in these Proceedings), and
other high-z IFS surveys.

Here we describe two wide-field IFS surveys targeting a distinct sample
of nearby objects: a) the PPAK IFS Nearby Galaxies Survey: PINGS (Rosales-Ortega
et al. \cite{pings}, focusing on spiral galaxies); and b) the VIMOS-VLT survey of
(Ultra)luminous Infrared Galaxies (U-LIRGs, Arribas et al. \cite{vimos}). We
present the observations and preliminary results on the chemical
abundance analysis, with an emphasis on the radial abundance gradients, the
2D distribution of the metal content in the galaxies, and differentials found
between extra-nuclear and nuclear abundances, based on S/N-optimised integrated
spectra.

\section{The PPAK IFS Nearby Galaxies Survey: PINGS \label{pings}}

The PPAK IFS Nearby Galaxies Survey (PINGS, Rosales-Ortega et al. \cite{pings})
consists in a wide-field 2D spectroscopic survey of a sample of nearby spiral galaxies
in the optical wavelength range, using the PMAS/PPAK
spectrograph
mounted at the 3.5m telescope of the Centro
Astron\'omico Hispano-Alem\'an at Calar Alto, Spain.
This project represents the first attempt to obtain continuous coverage
spectra of the whole surface of a galaxy in the nearby universe.
The observations consisted of Integral Field Unit (IFU) spectroscopic mosaics
for 17 galaxies within a maximum distance of 100 Mpc; the average distance
of the sample being 28 Mpc (for $H_0$ = 73 km\,s$^{-1}$\,Mpc$^{-1}$). The
sample includes normal, lopsided, interacting and barred spirals with a good
range of galactic properties and star-forming environments with multi-wavelength
public data (e.g. see Fig.~\ref{fig1}).
The spectroscopic mosaicking was acquired during a period of three years and the
final data set comprises more than 50\,000 individual spectra, covering in total
an observed area of nearly 80 arcmin$^2$, an observed surface without
precedents by an IFS study up to now. The observations are currently being supplemented
with broad band and narrow band imaging for those objects without public
available images in order to maximise the scientific and archival value of the
dataset.

\begin{figure}
\center
\includegraphics[height=6.8cm]{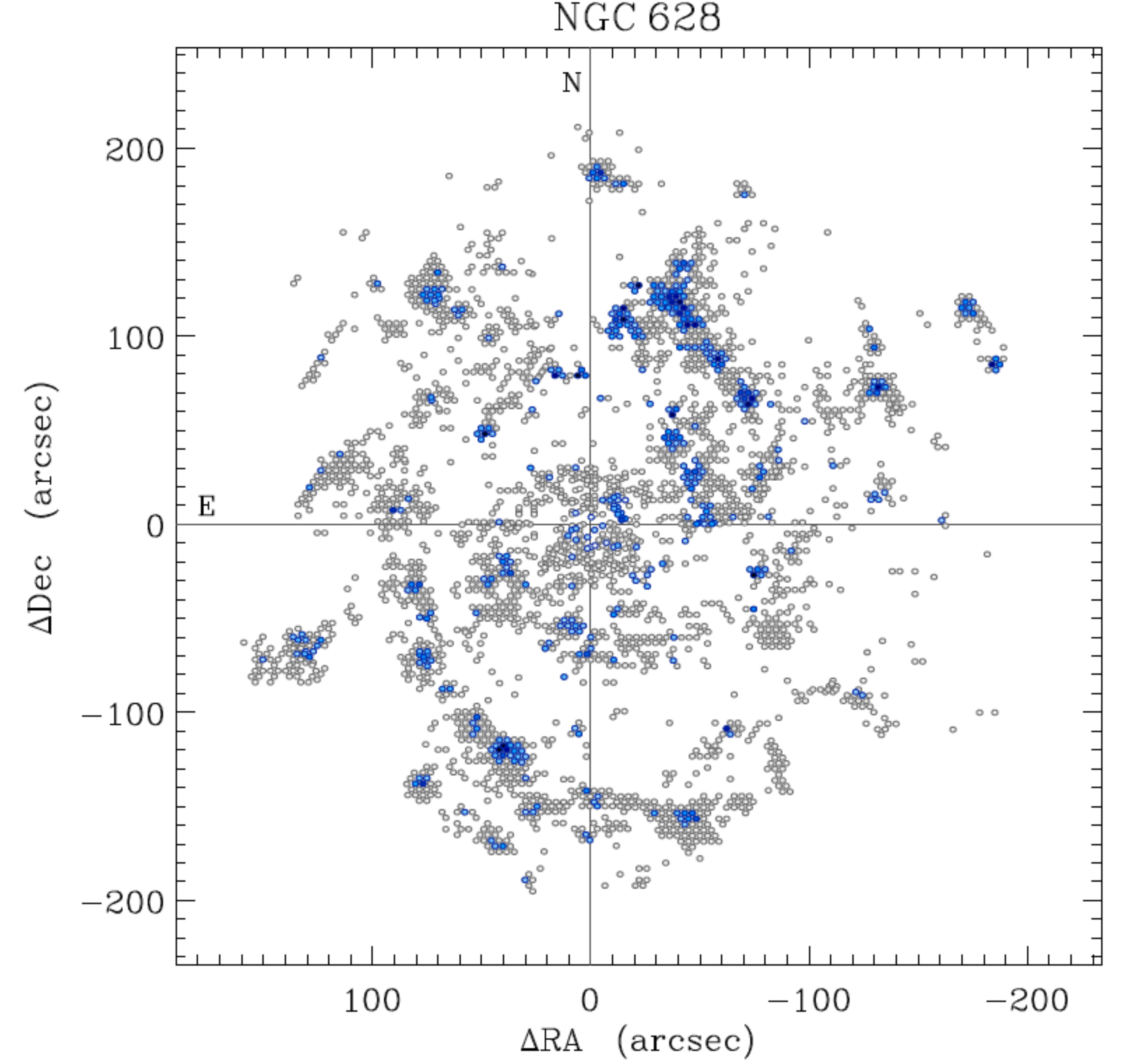} ~
\includegraphics[height=6.6cm]{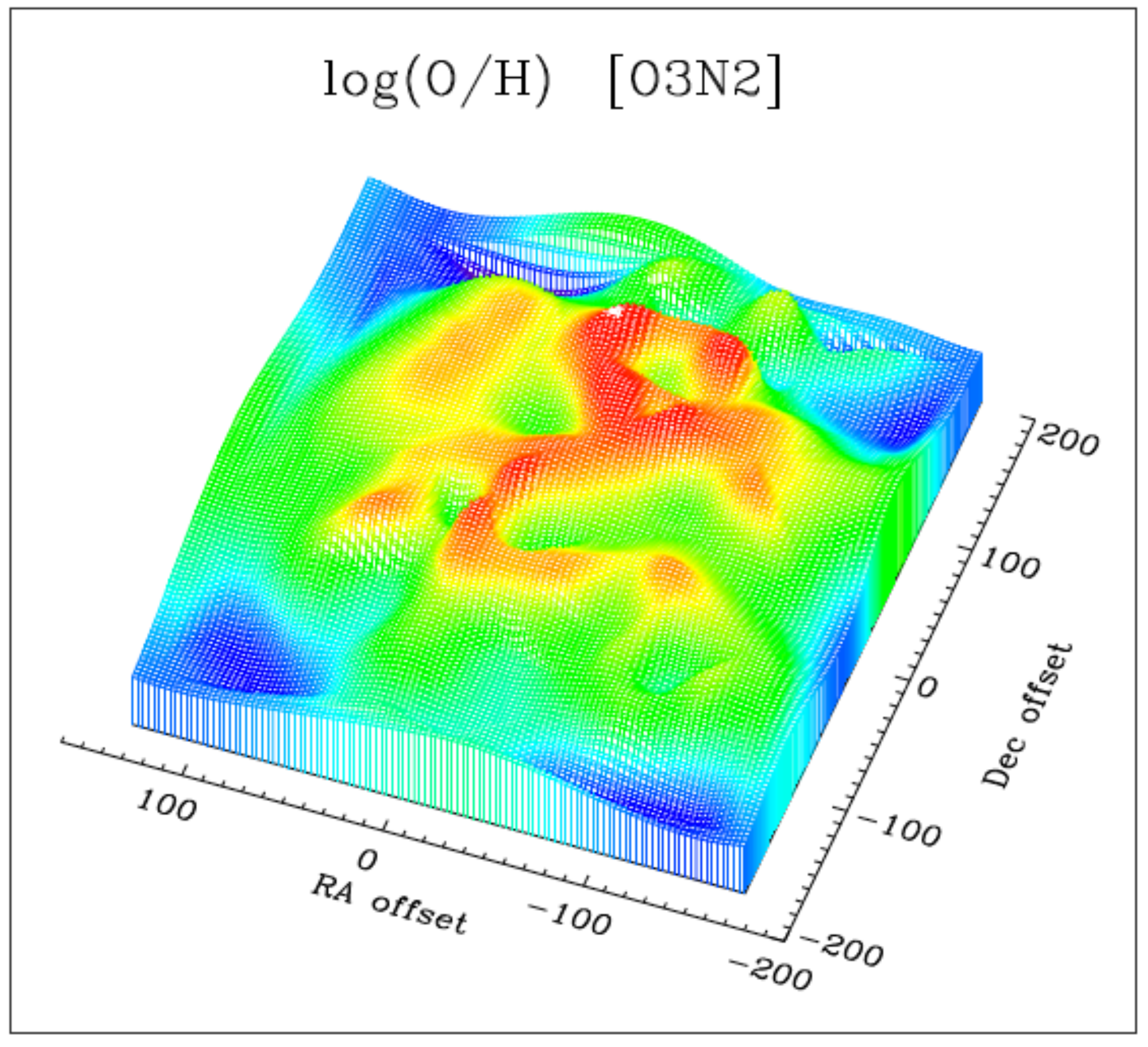} 
\caption{\footnotesize
  {\em Left}: Spatial map of the
  PINGS mosaic of NGC\,628 showing those fibres
  for which nebular emission with high S/N is detected (grey), and those with enough
  emission lines for a complete chemical abundance analysis (blue). In the
  latter case, the colour intensity of each fibre has been scaled to the flux
  intensity of H$\alpha$ for that particular spectrum.
  {\em Right}: 3D rendering of an interpolated grid of oxygen abundances derived
  from the IFS mosaic of NGC\,628, showing a surface of iso-abundance contours
  obtained using the O3N2 metallicity calibrator.
  \label{fig2}
}
\end{figure}

As part of the main data products, we have obtained for each IFS mosaic: a)
complete pixel-resolved maps of the emission-line abundances based on a suite of
strong-line diagnostics, incorporating absorption-corrected H$\alpha$, H$\beta$,
\oii, \oiii, \nii, and \sii\ line ratios (e.g. see Fig.~\ref{fig2}); b) local
nebular reddening estimates based on the Balmer decrement; c) measurements of
ionisation structure in \hh regions and diffuse ionized gas (DIG) using the well-known and most
updated forbidden-line diagnostics in the oxygen and nitrogen lines; d) rough
fits to the stellar age mix from the stellar spectra of the galaxy sample
(S\'anchez et al. \cite{n628}).

With this approach we intend to obtain relations between the 2D distribution of
gas metallicity and galaxy structure. The spatial study of abundance
distributions is enabling us to detect features not previously observed by
one-dimensional or single-aperture studies, i.e. inner and/or outer radial
gradients, non-axially symmetric abundance distributions, shoulders or local
perturbations of the abundance distribution, etc. 
The presence of these features would give us information about,
e.g. anisotropic gas flows across the galaxies, the presence of galaxy
environmental effects, or features currently predicted only by theory or
inferred by a handful of observations (e.g. Vorobyov \cite{voro}, Bresolin et
al. \cite{m83}). Furthermore, the 2D line-ratio maps are allowing us to study the
physical state and ionization mechanisms of the DIG.

Likewise, the nature of the scaling laws relating metallicity and star
formation to other fundamental galaxy properties like luminosity, mass or
surface brightness are being studied in detail.
In most cases these relations were derived for integrated (or nuclear)
properties of the galaxies, not for resolved structures. This aperture problem
seems particularly relevant for metallicity since chemical abundance gradients
appear to be present in most galaxies. The
spatially resolved information provided by PINGS will bring light in this
problem, allowing to properly study the metallicity scaling laws. This in turn
has implications for broader problems within astrophysics, such as an improved
understanding of how star formation proceeds at different masses, luminosities,
metallicities and environments that might be of relevance at earlier epochs, and
therefore, our confidence in the use of metallicity calibrators applied to
high-z objects will be considerably enhanced.

\begin{figure}
\center
\includegraphics[height=5.8cm]{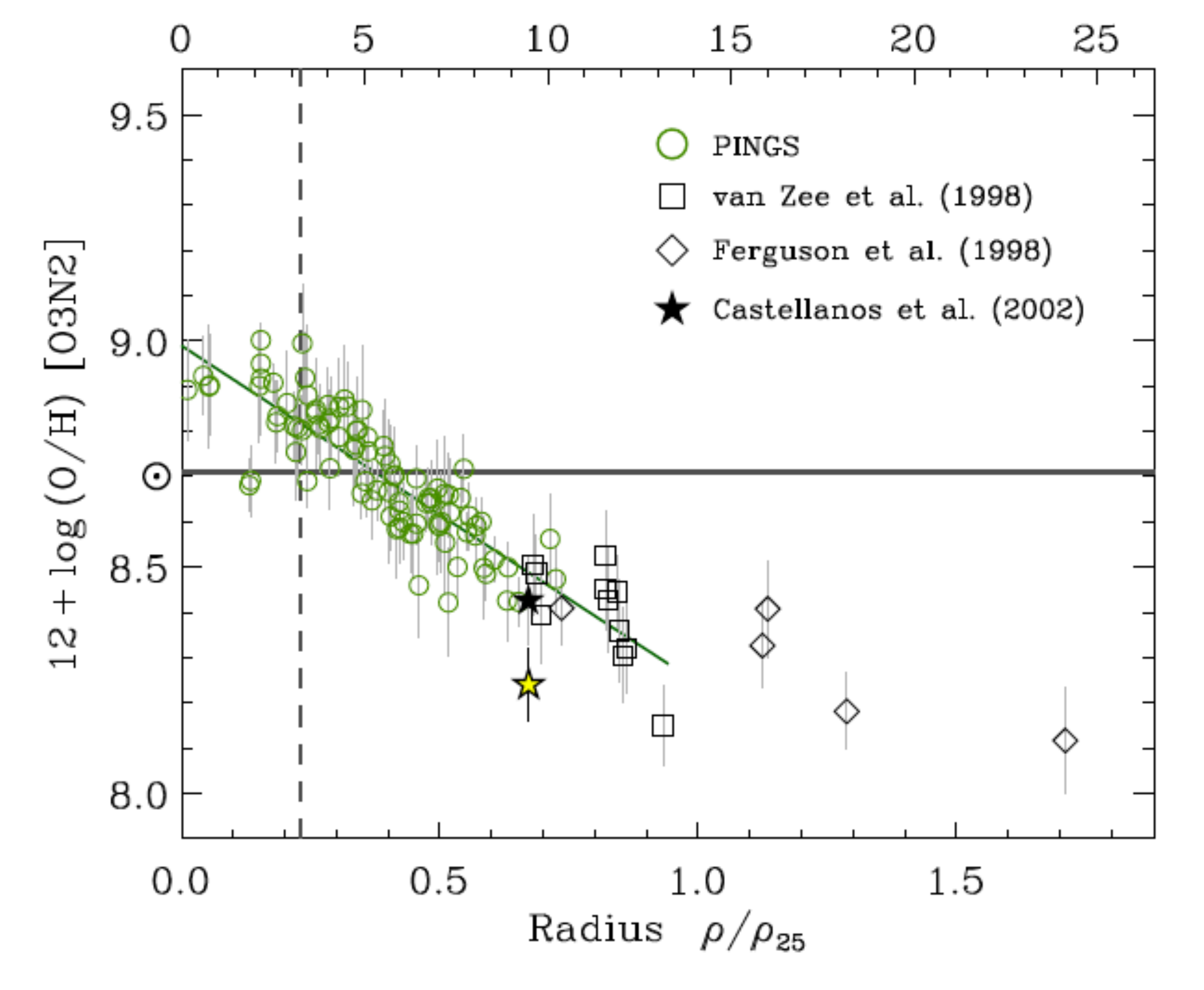} 
\includegraphics[height=5.5cm]{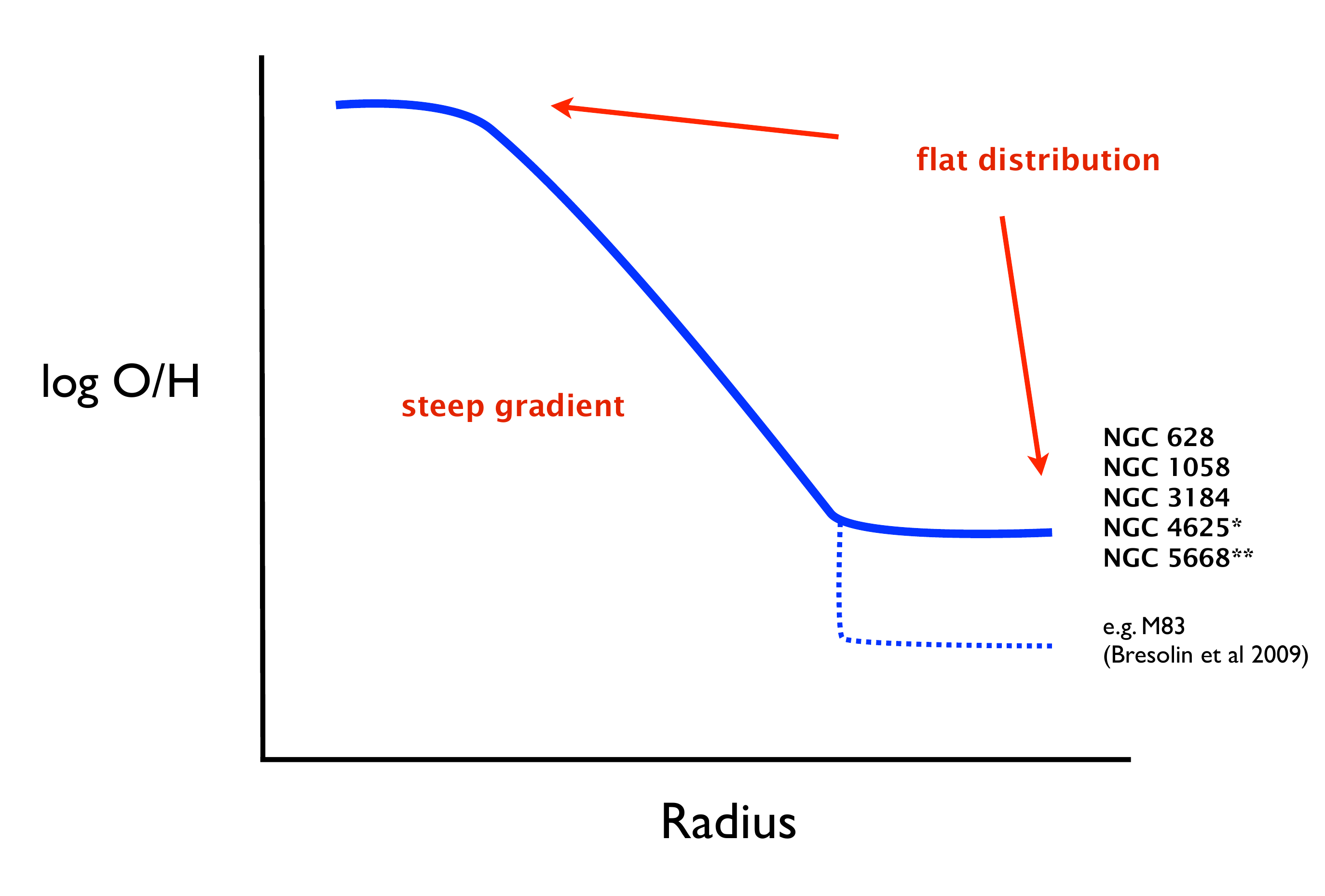} 
\caption{\footnotesize
  {\em Left}: Oxygen abundance vs. normalised radius (metallicity gradient) of NGC\,628 based on HII
  regions selected within the FoV of the PINGS mosaic (green), and from the
  literature (black symbols), derived using the O3N2 calibrator
  (Rosales-Ortega et al. \cite{n628ii}). 
  {\em Right}: Multi-modality of the abundance gradient inferred for a number of
  spiral galaxies in the PINGS sample (and in the literature), consistent with a
  flat-steep-flat behaviour with increasing galactocentric radii. In the case of M83, a
  discontinuity might also be present. 
  {\em Notes}: $^{\ast}$Goddard et al. \cite{n4625}; $^{\ast\ast}$Marino R. (PhD thesis).
  \label{fig3}
}
\end{figure}

Some of the preliminary results arising from these studies are the following:
a) we found compelling evidence suggesting that, measurements of emission lines of
classical \hh regions are not only aperture, but spatial dependent, and
therefore, the derived physical parameters and metallicity content may
significantly depend on the morphology of the region, on the extraction
aperture and on the S/N of the observed spectrum (Rosales-Ortega et
al. \cite{pings}, S\'anchez et al. \cite{n628}). b) We found observational
evidence of non-linear multi-modal abundance gradients in normal spiral
galaxies, consistent with a flattening in the innermost and outermost parts of
the galactic discs, with important implications in terms of the chemical
evolution of galaxies (Rosales-Ortega et al. \cite{n628ii}, see Fig.~\ref{fig3}).

Such multi-modal gradients have been theoretically predicted,
e.g. in the case of the inner plateau, this can be due to the influence of a
non-axisymmetric gravitational field of spiral density waves (Vorobyov
\cite{voro}), and in the case of the flattening at large galactocentric
distances, the effect can be reproduced by ``inside-out'' scenarios of galaxy
formation, in which the galaxy disk is built up via gas infall, and in which the
timescale for the formation of the disk increases with galactocentric distance 
(e.g. Bresolin et al. \cite{m83}, and references therein). Although previous
observations showed complex non-linear behaviour in the radial distribution of
oxygen for some spiral galaxies (e.g. Zaritsky et al. \cite{zar}), the nearly 2D
coverage of the PINGS mosaics allows to provide strong observational evidence
of these effects, especially in the innermost regions of the galaxies where no
spectroscopic studies were attempted before (e.g. NGC\,628, NGC\,3184, etc.).

In summary, most studies to date for obvious reasons have preferentially
targeted the nebular emission of the brightest and highest surface
brightness \hh regions, but by targeting virtually every \hh region within the 
galaxies discs, our IFS data is providing a much fuller and more objective
sampling of the \hh region population. In other words, similarly to SAURON for
early-type galaxies, PINGS is providing the most detailed knowledge of star
formation and gas chemistry across a late-type galaxy.

\section{The metallicity structure of low-z (Ultra)Luminous Infrared Galaxies (U)LIRGs \label{lirg}}

As mentioned before, understanding the evolution of the metal enrichment over
cosmic time is a core question in astrophysics.
Nearby (Ultra)Luminous Infrared Galaxies (LIRGs: $L_{IR} \equiv L[8-1000 \mu m]
= 10^{11-12}L_{\odot}$; ULIRGs: $L_{IR} > 10^{12}L_{\odot}$) are particularly
interesting local objects in that respect. They constitute the local counterpart
of massive, dusty high-z star-forming galaxies, for which higher S/N and better
linear resolutions can be easily achieved (e.g. P\'erez-Gonz\'alez et
al. \cite{perez}).
The observed properties of (U)LIRGs share many characteristics
with populations of star-forming galaxies at high-z (Frayer et
al. \cite{frayer}). Therefore, the local (U)LIRGs population provides the
opportunity to link the properties of those high-z sources with those we observe
in the nearby universe.

The ISM of interacting U-LIRGs is frequently found in a kinematically extreme
state, dominated by inflows, outflows, and turbulent motions (e.g. Veilleux et
al. \cite{vei}, Monreal-Ibero et al. \cite{monreal}),
which trigger an intense star formation, producing and redistributing metals at
a prodigious rate, altering significantly the chemical states of the progenitor
galaxies.
Previous works have measured the metal content in low-z (U)LIRGs by using
(long-slit or single-aperture) nuclear spectra for $\sim 100$ (U)LIRGs at
$\langle z \rangle \sim 0.1$ (Rupke et al. \cite{rupke}). These studies found
that (U)LIRGs are under-abundant by a factor of two on average, when compared
with local, non-interacting, emission line galaxies with modest star formation
and similar luminosity and mass (using the near-IR $L$ vs. $Z$ relation),
implying that local abundance scaling relations are not universal.

\begin{figure}
\center
\includegraphics[height=7.5cm]{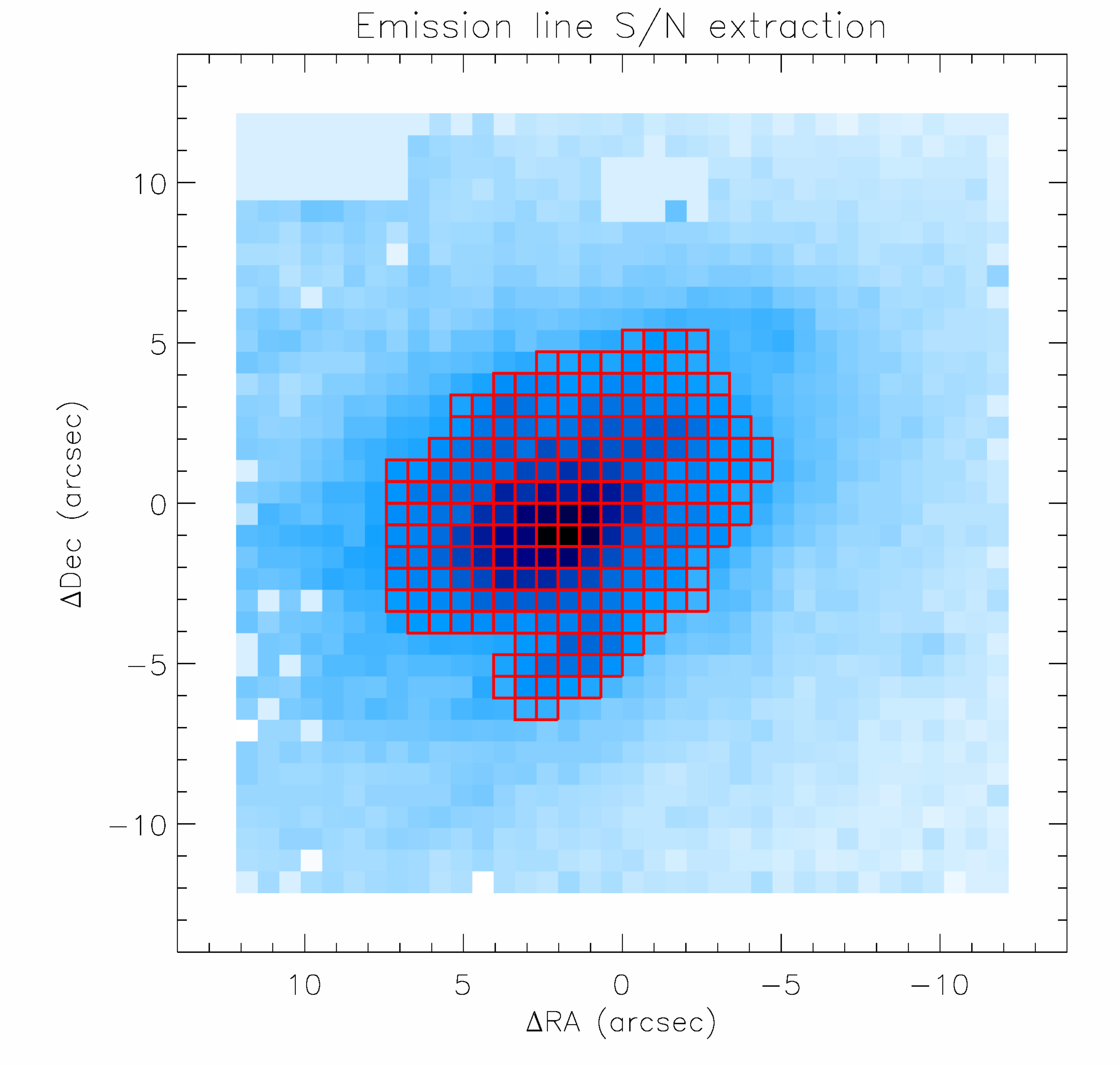} ~
\includegraphics[height=7.5cm]{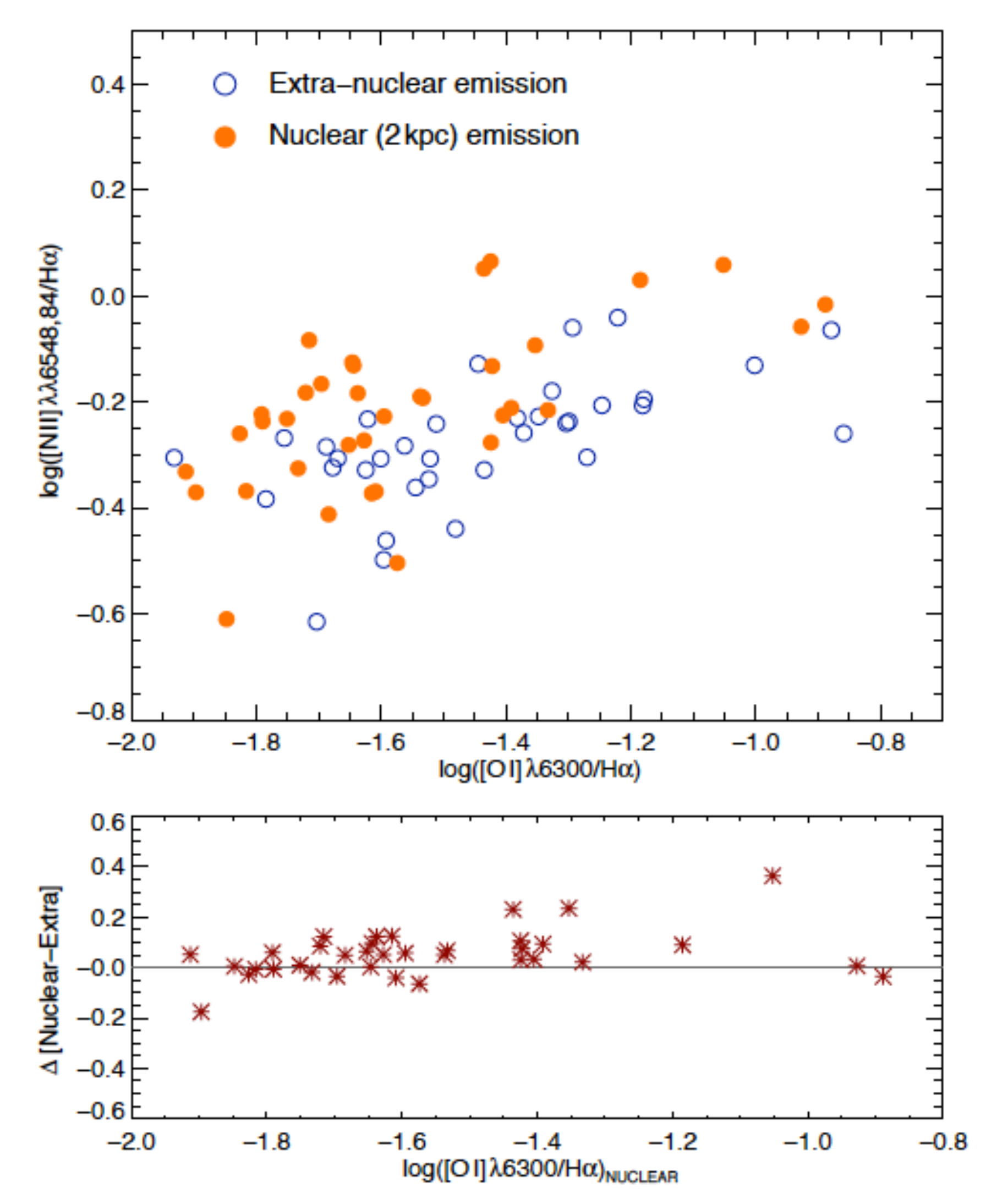} 
\caption{\footnotesize
  {\em Left}: VIMOS H$\alpha$ ``narrow band image'' of the LIRG IRAS F12115-4656
  after Arribas et al. \cite{vimos}. The red squares indicate the selected
  spaxels (after a S/N optimisation scheme) used to to obtain an integrated
  spectrum.
  {\em Right}: \nii/\ha\ vs. \oi/\ha\ diagnostic diagram for the integrated
  emission of the nuclear and extra-nuclear regions of the VIMOS-(U)LIRGs
  sample, showing the relative differences between the two populations.
  \label{fig4} 
}
\end{figure}

The SIRIUS project (Arribas et al. \cite{vimos}, Garc\'ia-Mar\'in et
al. \cite{gm}, Alonso-Herrero et al. \cite{alonso}), is a long-term programme
aimed at studying the internal structure and kinematics of a
representative sample of low-z (U)LIRGs, using a variety of optical and
near-infrared IFS facilities (INTEGRAL+WYFFOS \& PMAS in the northern hemisphere
and VIMOS-VLT \& SINFONI in the southern one). 
As part of this project, we are characterising the nebular properties of the VIMOS-VLT
sample, which contains a total of 42 systems with a mean redshift of 0.02, covering
different morphological classes (spirals, early interactions, advance mergers
and post-mergers). The purpose is to exploit the power of IFS observations by
obtaining S/N-optimised, velocity-field corrected, integrated spectra of the
sources, in order to study how the global nebular
chemical abundance and ionization properties compare to the resolved cases (by
applying empirical calibrations as those employed in high-z studies,
e.g. Pettini \& Pagel \cite{o3n2}, see Fig.~\ref{fig4}), as
well as trying to assess the contribution of the nuclear region and/or diffuse
medium to the integrated spectra (which is especially relevant in the studies of
similar non-resolved high-z systems). The results of these investigations will
hopefullt allow us to contribute in the study of the chemical evolution among IR
populations.

\section{Conclusions \label{fin}}

A good deal of our current understanding of the Universe is due to large
surveys, either in the imaging or in the spectroscopic domain. However, those
surveys conceived to provide spectral information (such as SDSS or zCOSMOS) are
generally limited to one spectrum per galaxy, often with aperture losses that
are difficult to control, leading to very different linear scales at different
redshifts. The SAURON, SIRIUS, VENGA, PINGS and CALIFA projects
represent a milestone in terms of the new generation IFS surveys, considering
that before their advent, the imaging-spectroscopy technique was not considered to be
implemented in large samples, i.e. in a {\em survey mode}.

Nearby galaxies offer the unique opportunity to study the SFH-ISM coupling on a
spatially resolved basis, over large dynamic ranges in gas density and pressure,
metallicity, dust content, and other physically relevant parameters of gas and
dust. Therefore, the ongoing (and the upcoming) wide-field, nearby IFS surveys
represent a powerful tool for studying effectively not only every \hh region
within the whole surface area of a galaxy (alleviating the aperture problem),
but also those regions where diffuse emission is present and --as very important
by-product-- a complete 2D picture of the underlying stellar populations of a
galaxy. These projects will provide the largest and most comprehensive IFS
survey of galaxies, which will allow the community to address several
fundamental issues in galactic structure and evolution. 
The future on this field looks very promising, and it will hopefully help to
bring new insights into the physical processes at play in star formation and
galaxy evolution.\\


\small  
%


%
\end{document}